\documentclass[12pt]{iopart}

\usepackage{latexsym}
\usepackage{mathrsfs}

\begin{document}

\maketitle

\title{Cherenkov radiation in a photon gas}

\author{Mattias Marklund, Gert Brodin, Lennart Stenflo, Padma K.\ Shukla}

\address{Department of Physics, Ume{\aa} University, SE--901 87
  Ume{\aa}, Sweden}

\date{\today}

\begin{abstract}
  It is well-known that a charged particle moving 
  with constant velocity in vacuum does not radiate. In a  
  medium the situation can be different. If the so called Cherenkov 
  condition is satisfied, i.e.\ the particle velocity exceeds the phase speed
  in the medium, the particle will radiate. 
 We show that 
  a charge moving with a constant velocity in a gas of photons emits  
  Cherenkov radiation, even in the gamma-ray regime, due to nonlinear
  quantum electrodynamic effects. 
  Our result is evaluated with respect to the radiation background
  in the early universe, and it is argued that the effect can be significant. 
\end{abstract}
\pacs{41.60.Bq, 98.80.-k}

\section*{}

In 1934, Cherenkov observed the type of radiation now bearing his name
\cite{Cerenkov}. His experimental result was explained by Tamm \& Frank 
\cite{Tamm-Frank}. In an isotropic 
dielectric medium, a charged particle in rectilinear motion  
satisfying the so called Cherenkov condition, i.e.\  its
velocity exceeds the (parallel) phase speed in the medium in which it 
moves, will thus radiate \cite{Chefranov}. The radiation chock-front, called the Cherenkov cone, is
analogous to the Mach cone formed as objects move with supersonic 
speeds through air. 
In quantum mechanical terms, the Cherenkov condition corresponds to energy
and momentum conservation. 
Cherenkov radiation has technological uses, e.g.\  in
determining particle velocities. 

Quantum electrodynamics (QED) predicts many phenomena with no classical
counterparts, such as the Casimir effect and elastic photon--photon scattering
\cite{Heisenberg-Euler,Weisskopf,Schwinger}.
While the former has been experimentally confirmed, the latter has still to
be detected \cite{Brodin-Marklund-Stenflo}. In addition, the effective field theory describing photon--photon 
scattering \cite{Schwinger} has been widely used to predict possible effects in, 
for example,  extreme magnetized objects, such as neutron stars and magnetars. 
One effect of a strong magnetic field is to down-shift the frequency of a 
test photon, the so called photon splitting \cite{Adler,Adler2}. 

In this Letter, we predict that a charged particle moving in an equilibrium
radiation gas will emit Cherenkov radiation. The possible importance
of this effect is then discussed for the cosmic radiation background.  
A related, but significantly different, study was presented by Dremin \cite{Dremin}.
The dispersion relation for electromagnetic waves in an isotropic
and homogeneous photon gas with refractive index $n$ 
is $\omega = kc/n$, where $n^2 = 1 + \delta$  
and \cite{Bialynicka-Birula,Marklund-Brodin-Stenflo}
\begin{equation}
  \delta = \frac{2b\alpha \mathscr{E}}{135 \pi \epsilon_0E_S^2} 
  \approx \frac{b\mathscr{E}}{4 \times 10^{29}\,\mathrm{J/m^3}} ,
\end{equation}
where $b$ is $8$ or $14$ depending on the photon polarization, 
$\alpha \approx 1/137$ is the fine structure constant, $\epsilon_0$ is the permittivity 
of vacuum, $\mathscr{E}$ is the energy 
density of the radiation gas, and $E_S \equiv m_e^2c^3/e\hbar \approx  10^{18} \, 
\mathrm{V/m}$ is the Schwinger field strength. Thus, the refractive index 
in this case is always larger than one, and a particle may
therefore have a speed $u$ exceeding the phase velocity in the
medium. The Cherenkov condition $u \geq c/n$ for emission of Cherenkov 
radiation can thus be satisfied. This condition can also be expressed in terms of the 
relativistic gamma factor $\gamma = (1 - u^2/c^2)^{-1/2}$, namely  
$\delta\gamma^2 \geq 1 $.  
We will here assume that a particle with charge $Ze$,  
satisfying the Cherenkov condition, moves through an equilibrium 
radiation gas.  
The energy loss at the frequency $\omega$ per unit length of the path of the 
charged particle is then \cite{Panofsky-Phillips}
\begin{equation}
  \frac{dU_{\omega}}{ds}d\omega = \frac{Z^2\alpha}{c}\frac{(\delta\gamma^2 - 1)}{(\gamma^2 - 1)}
  \hbar\omega \,d\omega ,
\end{equation}
and the number of quanta $N$ emitted per unit length along the particles path 
is 
\begin{equation}
  \frac{dN}{ds} d\omega= \frac{Z^2\alpha}{c} \frac{(\delta\gamma^2 - 1)}{(\gamma^2 - 1)} \, d\omega .
\end{equation} 
Since $\delta$ is normally much less than one, we need a large gamma factor to satisfy the 
Cherenkov condition. Moreover, the present theory of photon--photon scattering is only valid 
as long as $\omega \ll \omega_e = m_ec^2/\hbar \approx 8\times 10^{20} \,\mathrm{rad/s}$.
The Compton frequency $\omega_e$ acts as a cut-off in the integration of the 
formulas for the energy loss and number of quantas. Subsequently, for $\delta \gamma^2 =1$,
we have 
\begin{equation}
  U = N\hbar\omega_e, \quad \quad N = Z^2L\alpha\delta/\lambda_e ,
\end{equation}
respectively. Here $L$ is the distance traveled by the charge, and 
$\lambda_e = c/\omega_e$ is the Compton wavelength. 

At the present time, the cosmic microwave background has an energy
density of the order $\mathscr{E} \sim 10^{-15}\,\mathrm{J}/\mathrm{m}^3$, i.e.\
$\delta \sim 10^{-42}$, i.e. the gamma factor has to be 
$\gamma \geq 10^{21}$ for the Cherenkov condition to be satisfied.
Thus Cherenkov radiation is not likely to occur in todays radiation
background. In fact, it is well known that the cosmic rays contain non-thermal hadrons, of which some 
are protons, that can reach gamma factors $10^{11}$, but larger values are
improbable due to the GZK cut-off \cite{Greisen,Zatsepin-Kuzmin}. 
As a comparison, we may consider the situation at the time of
matter--radiation decoupling. 
Since $\mathscr{E}_{\mathrm{emitted}} = \mathscr{E}_{\mathrm{received}}(T/2.7)^4$, 
where the temperature $T$ is given in Kelvin, we have $\mathscr{E} \sim 10^{-2} \, \mathrm{J/m^3}$ 
at the time of decoupling ($T \approx 8000\, \mathrm{K}$), implying $\delta \sim 10^{-28}$. 
Thus, the limiting value on the gamma factor 
for the Cherenkov condition to be satisfied is $\gamma \geq 10^{14} - 10^{15}$,
still out of reach for high energy cosmic rays.
However, as we demonstrate below, the situation changes
drastically for earlier processes at even higher $T$. In particular we will focus on the era with
$10^9\,\mathrm{K} \leq T \leq 10^{11}\,\mathrm{K}$ when the reguired $\gamma$-factors range from
$\gamma \sim 10^4$ to $\gamma >3$.

The effect presented above is then naturally compared with inverse Compton
scattering. Setting $Z = 1$, the cross section for this scattering is $\sigma \approx \pi r_e^2m_e^2/M^2\gamma$, 
where $r_e$ the classical electron radius and $M$ is the charged particle mass. We thus 
obtain a collision frequency $\nu = c\mathscr{N}\sigma$, where $\mathscr{N}$ is 
the number density of the photons. Comparing this frequency with 
the frequency $\nu_{\mathrm{ch}} = (\gamma Mc)^{-1}dU/dt$, we note that fast 
particles are mainly scattered due to the Cherenkov effect when $\nu < \nu_{\mathrm{ch}}$, i.e.
\begin{equation}\label{eq:cherenkov}
  1 <  \frac{\delta}{\alpha\pi(m_e/M)\mathscr{N}\lambda_e^3} =  \frac{M}{m_e}\frac{T}{T_{\mathrm{ch}}}.
\end{equation}
Here $T$ is the temperature of the photon gas,  
$\mathscr{N} = [30\zeta(3)a/k_B\pi^4]T^3$, $\mathscr{E} = aT^4$, 
$k_B$ is the Boltzmann constant,  
$a = \pi^2k_B^4/15\hbar^3c^3 \approx 7.6\times 10^{-16}\,\mathrm{J/m^3K^4}$ 
and $T_{\mathrm{ch}}= 2025\zeta(3)m_ec^2/4\pi^3\alpha bk_B \approx 10^{12}\,\mathrm{K}$.
Thus, for a single fast proton to be scattered mainly due to the Cherenkov effect,  
we need $T > T_{\mathrm{ch}}\times 10^{-3} \sim 10^9 \, \mathrm{K}$, well 
within the limit of validity of the theory for photon--photon scattering.    
We note that at radiation gas temperatures around $10^{12}\,\mathrm{K}$ the 
quantum vacuum becomes truly nonlinear, and higher order QED effects should
be taken into account. 

For the early universe considered above,  
a moderately relativistic plasma is also present, which means that collective 
charged particle interactions can play a role. We take  
these plasma effects into account by introducing the plasma frequency $\omega_p$. 
The photon dispersion relation then reads $\omega
^{2}\approx k^{2}c^{2}(1-\delta )+\omega _{p}^{2}$. Thus the
Cherenkov condition is satisfied for charged particles with relativistic factors $\gamma \geq 1/%
\sqrt{\delta -\omega _{p}^{2}/k^{2}c^{2}}$. For the temperatures where Cherenkov
radiation starts to dominate over inverse Compton scattering, $T\sim
10^{9}-10^{10}\,\mathrm{K}$, we have  $\omega _{p}\sim 10^{15-16}\,\mathrm{rad/s}$,
and thus Cherenkov radiation is emitted in a broad band starting in the 
UV range, $\omega\sim 10^{17}\, \mathrm{rad/s}$, and continuing 
up to the Compton frequency $\sim 8\times 10^{20} \,\mathrm{rad/s}$. 

The Cherenkov radiation emitted during the era when $T \sim 10^9\,\mathrm{K}$
will be redshifted due to the cosmological expansion. Thus, the present value of the 
cut-off frequency will be approximately $2\times 10^{12}\,\mathrm{rad/s}$, i.e., in
the short wavelength range of the microwave spectrum. However, we do not expect 
direct detection of this radiation in the present universe, since the process is
expected to be of importance long before the time of radiation decoupling. Still, there are 
possible important observational implications due to the Cherenkov mechanism
presented here. As shown by the inequality (\ref{eq:cherenkov}), the effect will be more pronounced
for massive particles with a given gamma factor, and protons are therefore expected to 
be more constrained than electrons by the QED Cherenkov emission. In particular,
(\ref{eq:cherenkov}) puts stronger limits than Compton scattering for  
supra-thermal protons observed today to be relics of the early universe.
In fact, it seems rather unlikely, given the inequality (\ref{eq:cherenkov}), that
such protons could survive during the $T = 10^9 - 10^{10}\,\mathrm{K}$ era.         

Thus, it cannot be excluded that the QED Cherenkov effect in the early universe can be 
significant even for todays observations.

\end{document}